# Phase Matching of High-Order Harmonics in Hollow Waveguides

Charles G. Durfee III, Andy R. Rundquist, Sterling Backus, Catherine Herne,

Margaret M. Murnane, and Henry C. Kapteyn

*Center for Ultrafast Optical Science*

*University of Michigan*

*2200 Bonisteel Blvd.*

*Ann Arbor, MI 48109-2099*

*Ph. (734) 763-4875; FAX: (734) 736-4876; E-mail: charlesd@eecs.umich.edu*

**Abstract**

We investigate the case of phase-matched high-harmonic generation in a gas-filled capillary waveguide, comparing in detail theory with experiment. We observe three different regimes of phase matching: one where atomic dispersion balances waveguide dispersion, another corresponding to non-collinear Cerenkov phase-matching, and a third where atomic dispersion and plasma dispersion balance. The role of atomic dispersion is demonstrated by studying the dependence of the harmonic signal for several gases. We also predict and provide preliminary evidence of a regime where phase-matching occurs only at specific fractional ionization levels, leading to an output signal that is sensitive to the absolute phase of the carrier wave.

In the process of high-harmonic generation (HHG), an atom illuminated by light of ionizing intensity radiates coherent harmonics of the incident laser that extend into the soft-x-ray region of the spectrum.[1-9] From a fundamental physics point of view, HHG explores the boundaries between classical and quantum behavior. It is also of interest as a practical coherent ultrafast radiation source in the VUV and soft-x-ray region of the spectrum. Finally, HHG is also believed to be a viable method for producing sub-femtosecond-duration light pulses.[8, 10, 11] In its application as a coherent light source, it is desirable to obtain the highest possible conversion efficiency of laser light to x-rays. This conversion efficiency depends both on the fundamental laser-atom interaction and on the buildup of the signal electromagnetic wave over an extended interaction length.[2, 12] In the former case, there is a tradeoff between the magnitude of the susceptibility and the highest, "cutoff" harmonic order, which is related to the ionization potential of the atom. Molecular and cluster species have also been investigated for high-harmonic generation as a means of obtaining even higher susceptibilities.[13] Macroscopic propagation effects play a dramatic role in the observed signal intensity. In past work, HHG was implemented by focusing an intense femtosecond-duration light pulse into a gas cell or jet.[2, 12, 14] This work has demonstrated the importance of the relative phase shift of the various interacting waves as they pass through the focus (i.e. the Guoy phase shift). While some enhancement comes from positioning the interaction region after the focus of the beam[15, 16], the divergence of the laser beam and the nonlinear dependence of the propagation phase on the axial distance in this case inherently limit the interaction length or require a large confocal parameter (and large laser energy) to implement efficiently.

In recent work, we demonstrated that this limitation could be circumvented by generating high-harmonics with sub-millijoule energy ultrashort light pulses guided in a gas-filled capillary waveguide.[17] As will be discussed below, the plane-wave geometry of the mode makes it possible to balance the anomalous dispersion of the neutral gas with the normal dispersion of the waveguide and the free electrons, over an extended interaction length.[18] By adjusting the gas pressure within the waveguide to tune the phase-velocity of the fundamental, phase-matched conversion of 800nm light to wavelengths around 25nm was demonstrated in argon. In this paper, we report on several new experimental results along with a model that sheds light on the physical mechanisms involved in the phase matching. Comparison of measurement and theory demonstrates the dependence of the optimum phase matching pressure on gas species, the effects of absorption on the harmonic signal, and reveals that confinement of the fundamental mode in the waveguide allows higher levels of ionization to be compensated than would be expected based on a 1-D plane-wave analysis. It has not been recognized previously that even in the absence of a waveguide, neutral-gas dispersion can play an important role in the observed HHG conversion efficiency. We show three limiting regimes of phase matching: a balance between the neutral gas and the waveguide dispersion, the neutral gas and the plasma dispersion, and Cerenkov (noncollinear) phase matching. In the second of these regimes, we predict that phase matching occurs only at specific ionization fractions, leading to an output signal that is sensitive to the absolute phase of the carrier wave with respect to the pulse envelope.

A beam confined as a mode in a waveguide has a flat wavefront, enabling the scaling of phase matching to long interaction length[18]. For light traveling in the fundamental mode of a step-index waveguide of radius $a$, and filled with a medium of refractive index $n$, the longitudinal propagation constant, $k$, is given by $k^2 = n^2 k_0^2 - (u_{11}/a)^2$. Here $k_0$ is the vacuum wavenumber and $u_{11} = 2.405$ is the first zero of the Bessel function $J_0$.[19] The refractive index of the gas can be written as $n = 1 + P[(1-\eta)\delta(\lambda) - \eta N_{atm} r_e \lambda^2/2\pi + (1-\eta)n_2 I]$, where $P$, $\eta$, $N_{atm}$, $\delta(\lambda)$, $r_e$, $n_2$, and $I$ represent the pressure in atmospheres, the ionization fraction, the number density at one atmosphere, the neutral gas dispersion, the classical electron radius, the nonlinear refractive index and the pulse intensity, respectively. This gives the following propagation constant –

$$k \approx \frac{2\pi}{\lambda} + \frac{2\pi P(1-\eta)\delta(\lambda)}{\lambda} + (1-\eta)n_2 I - P\eta N_{atm} r_e \lambda - \frac{u_{11}^2 \lambda}{4\pi a^2} \quad (1)$$

where the terms on the right correspond to the contributions from vacuum, neutral gas dispersion, the nonlinear refractive index and waveguide dispersion, respectively. For the capillary lengths considered here, the harmonic light has a confocal parameter much longer than the capillary length, and effectively does not encounter the capillary walls. We therefore neglect the x-ray modal contribution.

Since ionization is localized near the optical axis, only a portion of the fundamental mode propagates through the plasma. Provided the plasma density is sufficiently small, the shape of the guided mode field is not greatly affected. In this regime, the radial variations in the refractive index may be treated as a perturbation, in a similar approach to that taken with self-phase modulation in optical fibers[20]. We calculate a new effective index by averaging $n(r)$ over the fundamental mode. This method is less accurate at high density and/or strong ionization, where a time-dependent beam propagation method must be used. This modal averaging reduces the effect of the plasma density by as much as a factor of five, and allows phase matching of higher-order harmonics than could normally occur with plane waves. We retain the nonlinear index in the calculations, but its effect is small: in argon, $n_2 I$ is only about 7% of the linear component $N_a \delta$ at an intensity of $2 \times 10^{14}$ W/cm² (using a value of $n_2 = 9.8 \times 10^{-24}$ cm²/W at one bar[21]—a possible overestimate by ~4x.[22]).

The phase mismatch for $q^{th}$ harmonic generation in a medium with a refractive index of $\eta$ is $\Delta k = k_q - qk_0 = qk_0(\eta(\lambda_q) - \eta(\lambda_0))$, or

$$\Delta k \approx q\frac{u_{11}^2 \lambda_0}{4\pi a^2} + N_e r_e (q\lambda_0 - \lambda_q) - \frac{2\pi N_a}{\lambda_q}(\delta(\lambda_0) - \delta(\lambda_q)), \tag{2}$$

where the nonlinear index term has been omitted. The gas dispersion is at photon energies greater than the ionization potential (i.e. for the harmonics). For photon energies $\varepsilon_{phot}$ lower than that of the first excited state, the contribution $N_a\delta$ to the refractive index is positive ($\approx +3\times10^{-4}$, for 1 bar of Ar at $\lambda_0 = 0.8\mu m$) while for $\varepsilon_{phot} > I_P$, it is negative ($\approx -1\times10^{-4}$, for $\lambda_q = \lambda_0/27$). By tuning the pressure, this anomalous dispersion is balanced with the normal dispersion of the waveguide and the free electrons to match the phase velocities of the fundamental and the harmonic light.

An important consideration in the phase matched generation of light with $\varepsilon_{phot} > I_P$ is absorption of the signal. Argon has a transmission window over the range 40-100 eV. In the presence of absorption, the equation for the growth of the signal wave can be written as

$$\frac{dE}{dz} \approx -\alpha E + iN_a \chi_{eff}^{(q)} E_0^s e^{-i\Delta kz}, \tag{3}$$

where $\alpha$ is the field absorption coefficient, and the nonlinear source term is modeled simply as having an effective field dependence of order $s\sim 5$ [2]. For a constant axial density profile, this results in a solution -

$$|E|^2 \approx N_a^2 |\chi_{eff}^{(q)} E_0^s|^2 \left(\frac{1 + e^{-2\alpha L} - 2e^{-\alpha L}\cos\Delta kL}{\alpha^2 + \Delta k^2}\right), \tag{4}$$

which reduces to the familiar $sinc^2(\Delta kL/2)$ dependence for $\alpha = 0$. The important physical consequence of Eqn. 4 is that the signal strength is determined by the shorter of the coherence length and the absorption depth, $l_{abs} = 1/2\alpha$. The strong absorption of gases for photon energies greater than the ionization potential therefore dominates the yield that can be obtained at different harmonics.[23, 24] This limitation can be circumvented by the use of short pump pulses (presented here), the use of wave-mixing schemes[25], or more complicated geometries[26], by operating in the transparency regions of the various gases.

Our experimental setup is similar to that described in our earlier work.[17] Briefly, laser pulses from a Ti:sapphire multi-pass amplifier (<4.5 mJ, 20 fs, 800 nm) [27] were focused into a 150 μm inner-diameter capillary waveguide. Mode matching of the beam into the lowest spatial mode was critical in these experiments. The position of the focal spot was optimized with a 3-axis translation stage on the lens, while translating the beam before the lens with a parallel mirror pair controlled the angle of incidence. An input iris was used to fine-tune the focal spot diameter to match that of the lowest order mode of the waveguide (66% of the capillary diameter). A three-segment capillary tube (6.3 mm outer diameter) was used for these experiments, which allowed us to produce a constant pressure region within the center section, while the end sections exited to vacuum.[17, 23] The total capillary length was 6.4 cm, with a central section length of 3 cm. The harmonic spectrum was dispersed with an imaging grazing incidence spectrometer (Hettrick HiREFs) and detected with a microchannel plate detector coupled to a phosphor screen. A thin aluminum filter (0.2 μm) was placed before the detector, which eliminated the fundamental and transmitted only the 11[th] - 45[th] harmonics (17 – 70 eV).

We performed a series of measurements that explore different regimes of phase matching. In the first regime, there is very little ionization and phase matching occurs primarily as a balance of the dispersion of the neutral atoms and the waveguide. This occurs at low intensity where there is little ionization. Figure 1a shows the dependence of the harmonic signal (19[th], 21[st], 23[rd]) on the pressure of argon in the center capillary section, at an intensity of $1.2 \times 10^{14}$ W/cm$^2$ (140µJ, 23 fs). At this intensity, the cutoff harmonic is the 25$^{th}$, and the ionization fraction at the peak of the pulse is 0.5%. The curves show two peaks. The low-pressure peak results from conversion within the central, constant density section of the cell. As there is very little ionization, the optimum pressure is at ~15 Torr, as predicted by theory. The high pressure peaks in Fig. 1(a) result from conversion within the last section of the capillary cell. Because of the pressure gradient between the two ends of this section, the pressure immediately drops (by ~3x) in this section as the gas accelerates into it[28]. The signal from this section is optimized at cell pressures that give the optimum pressure of ~15-20 Torr somewhere within the gradient, yielding a pressure peak that is shifted to higher pressure and broadened.

Figure 1(b) illustrates the important role of absorption. For harmonics 19-23, the signal is strongly absorbed by the neutral gas, as shown by the dashed curve. As the harmonic order increases, the gas transmission increases, giving rise to a stronger signal from the center section, where the constant pressure allows a phase-matched signal to build up over a longer distance. Clearly, relatively high transmission within the neutral gas is essential for high conversion efficiency. An increase in the incident intensity allows the generation of harmonics within the transmission window of argon, as the data shows in Fig. 1(b) (solid), where the peak intensity was $1.9 \times 10^{14}$ W/cm$^2$. This data sharply contrasts to the usual plateau of harmonics, and shows clearly that the phase matching allows the signal strength to be limited by the absorption depth.

We further explored the interplay of the neutral and plasma dispersion by measuring the pressure dependence of the harmonic signal for several gases – xenon, krypton, argon and hydrogen (see Fig. 2(a), respective harmonic orders 23, 25, 29, 25). The dependence of the phase matched condition on the atomic dispersion is illustrated clearly here: the more dispersive the gas, the lower the optimum pressure for phase-matching, since a lower gas density is required to match the dispersion of the waveguide and free electrons generated from ionization. The higher fractional ionization here places these experiments in a transitional regime where the waveguide and plasma dispersion are equally important. It can easily be seen from Eqn. 3 that as $N_e$ increases, the atomic density must also be increased to maintain phase matching. For low ionization fractions $\eta$, this may be accomplished by increasing the gas density. The temporal dependence of the ionization level broadens and increases the phase-matching pressure peak.

We performed calculations to accurately model this transitional regime. In these calculations, we first compute the fractional ionization profile $\eta(r, t)$ using the ADK tunneling ionization rates[29], and then calculate the effective refractive index for each temporal slice of the fundamental. The harmonic light is generated and propagates in the ionized regions, so to a first approximation it sees the full level of ionization. The calculated pressure dependence of the yield is shown in Fig. 2(b). There is very good agreement between the width and locations of the pressure peaks to that found experimentally. The only fitting required was to adjust the incident intensity downward (by 15-30%) so that ionization did not dominate the phase matching. This would be experimentally consistent with a slight defocusing of the guided beam. Other uncertainties in the calculation are the exact refractive index of the harmonic and the ionization rate. We estimate that the ionization level at the peak of the pulse was ~6% (Xe and Kr), ~8% (Ar), and 2.6% (H).

The guided wave geometry provides for another very interesting regime of phase matching. At still higher intensity, the nature of the phase matching is dominated by the plasma dispersion. There exists a critical ionization fraction, $\eta_{cr}$, dependent on the gas-species, for which the dispersion of the neutrals is balanced by the plasma dispersion [23]. For plane-wave propagation in Ar and Xe, $\eta_{cr}$ = 4.8% and 11%, respectively. However, the averaging effect of the mode across the radial ionization profile allows this number to be increased significantly.

For intensities capable of producing the critical ionization fraction $\eta_{cr}$ at which there is a balance between plasma and atomic dispersion, the optimum yield is extremely sensitive to both the pressure and the ionization level. The rapid variation of the instantaneous intensity that results from the oscillations at the carrier frequency leads to incremental "step" increases in $\eta$ at each half-cycle of the field. As $\eta$ approaches $\eta_{cr}$, each of these steps should have a distinct optimum pressure (see Fig. 3(a)). Note that the location of these pressure peaks should be sensitive to the absolute phase of the carrier to the pulse envelope. Experimentally, the lack of control over this absolute phase leads to rapid fluctuations in the yield at high pressures (see Fig. 3(b)). These absolute phase effects may provide a method of generating efficient attosecond duration XUV pulses from the guided wave geometry, because the phase-matching is optimized only for a small time period within a single cycle of the driving field. We note that optimum phase matching is limited to those harmonics that can be generated with ionization levels $< \eta_{cr}$. Since atoms may experience higher intensity at a given level of ionization when irradiated with shorter pulses,[6, 30] a short pulse duration is critical in attaining high-order phase-matched harmonics.

By monitoring the output profile of the harmonic light, we observed a third regime of phase matching. The spatial profile of the phase-matched harmonic emission was measured by placing a microchannel plate 0.68 m away from the exit of the capillary. An aluminum filter was used to block the transmission of the fundamental. At low pressure, the spatial profile of the harmonic emission is somewhat large and diffuse (Fig 4a, 10 Torr). However, as the cell pressure is increased (Fig. 4b, 20 Torr), the spot size of the harmonic emission decreases until, at the phase-matched pressure (Fig. 4c, 45 Torr), a well defined output mode is observed. At still higher pressure, the signal strength decreases because of phase mismatch, but the output spot size no longer changes significantly. This evolution in spatial mode of the XUV light at low pressure results from noncollinear phase matching. At low pressure, the phase mismatch is dominated by the normally dispersive waveguide and plasma terms. Such a positive phase mismatch can be compensated if the harmonic emission occurs at an angle $\theta$ to the fundamental. In this case, the phase mismatch can be written as $\Delta k = k_h \cos\theta - qk_f$. This type of phase matching is sometimes referred to as Cerenkov phase matching,[31] since the phase velocity of the nonlinear polarization travels faster than the phase velocity of the harmonic in the medium. The maximum angle of emission $\theta_{max}$ corresponds to the highest phase mismatch, in this case the highest ionization fraction, $\eta$. Measured values of $\theta_{max}$ are plotted vs. pressure in Fig. 4d, together with the calculated curve. Here we use $\eta = 11\%$, consistent with our other calculations. The agreement at low pressures is good— as the phase mismatch approaches zero, the output beam is limited in size by the diffraction-limited divergence of the mode. At higher pressures, the beam approximately maintains this divergence but the signal strength decreases. Here the phase mismatch is negative for high pressures and the phase mismatch would not be decreased for emission at large angles. While angular emission of HHG has been observed previously in a gas jet experiment[32], we show here for the first time the transition from off-axis to axial phase-matching. Note that Cerenkov phase matching is not as efficient as phase matching in the forward direction, since the signal can coherently build up over a comparatively short distance before it leaves the source region.

In conclusion, we have explored several new regimes of phase matching for guided-wave high-order harmonic generation. Through careful comparison of our measurements with a simple propagation model, we have demonstrated the influence of the waveguide and gas dispersion, ionization, carrier phase, and geometry on phase matching. By controlling the gas species, waveguide diameter, ionization level, and gas pressure, phase matching was achieved over several cm lengths in Xe, Kr, Ar and H gases. Harmonic orders 19 – 47 have been phase-matched, corresponding to an energy range of 29 – 72 eV. At the highest orders, the efficiency is limited by the strong ionization. The most simple and efficient region to achieve phase matching corresponds to a balance between the dispersion due to the waveguide and plasma terms, and that due to the neutral gas atoms. This simple phase matching scheme is possible only for levels of the total fractional ionization that are sufficiently low that the plasma index of refraction does not dominate the phase mismatch. The fact that phase matching occurs below a particular level of ionization is not an overriding limitation: the ultrashort pulse duration (~20fs) allows atoms to experience high levels of intensity prior to ionization, thereby making possible the generation of harmonics at energies within the transmission window of the gas.[6, 9] Moreover, the harmonic emission is predicted to occur over an extremely short time-period of ~5fs.[8] Finally, we also predict a new regime in which phase-matching at a given pressure occurs only at specific ionization levels, leading to an output signal that is sensitive to the absolute phase of the carrier wave. This may allow very precise control over the temporal profile of the generated HHG signal.

The authors gratefully acknowledge funding for this work from the National Science Foundation.

**References**

[1] A. McPherson, G. Gibson, H. Jara, *et al*., J. Opt. Soc. Am. B **4**, 595 (1987).

[2] A. L'Huillier, K. J. Schafer, and K. C. Kulander, Phys. Rev. Lett. **66**, 2200 (1991).

[3] M. Lewenstein, P. Balcou, M. Y. Ivanov, *et al*., Phys. Rev. A **49**, 2117 (1993).

[4] K. C. Kulander, K. J. Schafer, and J. L. Krause, in *NATO 3rd Conference on Super Intense Laser-Atom Physics*, Han-sur-Lesse, Belgium, 1993).

[5] J. J. Macklin, J. D. Kmetec, and C. L. Gordon, III, Phys. Rev. Lett. **70**, 766 (1993).

[6] J. Zhou, J. Peatross, M. M. Murnane, *et al*., Phys. Rev. Lett. **76**, 752 (1996).

[7] I. P. Christov, J. Zhou, J. Peatross, *et al*., Phys. Rev. Lett. **77**, 1743 (1996).

[8] I. P. Christov, M. M. Murnane, and H. C. Kapteyn, Phys. Rev. Lett. **78**, 1251 (1997).

[9] Z. Chang, A. Rundquist, H. Wang, *et al*., Phys. Rev. Lett. **79**, 2967 (1997).

[10] P. B. Corkum, N. H. Burnett, and M. Y. Ivanov, Opt. Lett. **19**, 1870 (1994).

[11] K. Kulander, in *DAMOP*, 1996).

[12] A. L'Huillier, X. Li, and L. Lompre, J. Opt. Soc. Am. B **7**, 527 (1990).

[13] T. D. Donnelly, T. Ditmire, K. Neumann, *et al*., Phys. Rev. Lett. **76**, 2472 (1996).

[14] M. B. Gaarde, P. Antoine, A. L'Huillier, *et al*., Phys. Rev. A **57**, 4553 (1998).

[15] A. L'Huillier and P. Balcou, Phys. Rev. Lett. **70**, 774 (1993).

[16] P. Salieres, A. L'Huillier, and M. Lewenstein, Phys. Rev. Lett. **74**, 3776 (1995).

[17] A. Rundquist, C. G. Durfee, Z. Chang, *et al*., Science **280**, 1412 (1998).

[18] C. G. Durfee, S. Backus, M. M. Murnane, *et al*., Opt. Lett. **22**, 1565 (1997).

[19] E. A. J. Marcateli and R. A. Schmeltzer, Bell Syst. Tech. J. **43**, 1783 (1964).

[20]     G. P. Agrawal, *Nonlinear Fiber Optics* (Academic Press, 2nd Ed., San Diego, 1995).

[21]     H. J. Lehmeier, W. Leupacher, and A. Penzkofer, Opt. Commun. **56**, 67 (1985).

[22]     M. Nisoli, S. D. Silvestri, O. Svelto, *et al*., Opt. Lett. **22**, 522 (1997).

[23]     A. Rundquist, PhD thesis (Washington State University, Pullman, 1998), p. 185.

[24]     E. Constant, D. Garzella, P. Breger, *et al*., Phys. Rev. Lett. **82**, 1668 (1999).

[25]     P. L. Shkolnikov, A. E. Kaplan, and A. Lago, Opt. Lett. **18**, 1700 (1993).

[26]     I. P. Christov, H. C. Kapteyn, and M. M. Murnane, Opt. Expr. **3**, 360 (1998).

[27]     S. Backus, *et al*., Opt. Lett. **22**, 1256 (1997).

[28]     D. J. Santeler, J. of Vac. Sci. and Tech. A **4**, 348 (1986).

[29]     M. V. Ammosov, N. B. Delone, and V. P. Krainov, Sov. Phys. JETP **64**, 1191 (1986).

[30]     I. P. Christov, J. P. Zhou, J. Peatross, *et al*., Phys. Rev. Lett. **77**, 1743 (1996).

[31]     G. Leo, R. R. Drenten, and M. J. Jongerius, IEEE J. Quant. Elect. **28**, 534 (1992).

[32]     P. Salieres, T. Ditmire, M. Perry, *et al*., J. Phys. B **29**, 4771 (1996).

**Figure captions**

1. a) Pressure dependence of the signal for harmonics 19,21,23 generated in argon at a peak intensity of $1.2 \times 10^{14}$ W/cm$^2$. b) Harmonic spectrum at $1.8 \times 10^{14}$ W/cm$^2$ (solid) and the calculated transmission of 5mm neutral argon at 30 Torr.
2. Measured (a) and calculated (b) pressure dependence of the harmonic yield for several gases. In order of increasing optimum pressure, the curves correspond to xenon, krypton, argon and hydrogen.
3. a) Calculated pressure dependence of the yield for high intensity ($2.2 \times 10^{14}$ W/cm$^2$) in argon. b) Measured signal at an incident intensity of $4 \times 10^{14}$ W/cm$^2$.
4. a-c) Spatial profiles of the harmonic yield at 10, 20 and 45 Torr. The dark grid results from the mesh used to support the 0.2mm Al filter (70 lines/in). d) Maximum emission angle for different pressures (points); calculated emission angle for a peak ionization level of 11% based on Cerenkov analysis.

Figure 1 a-c

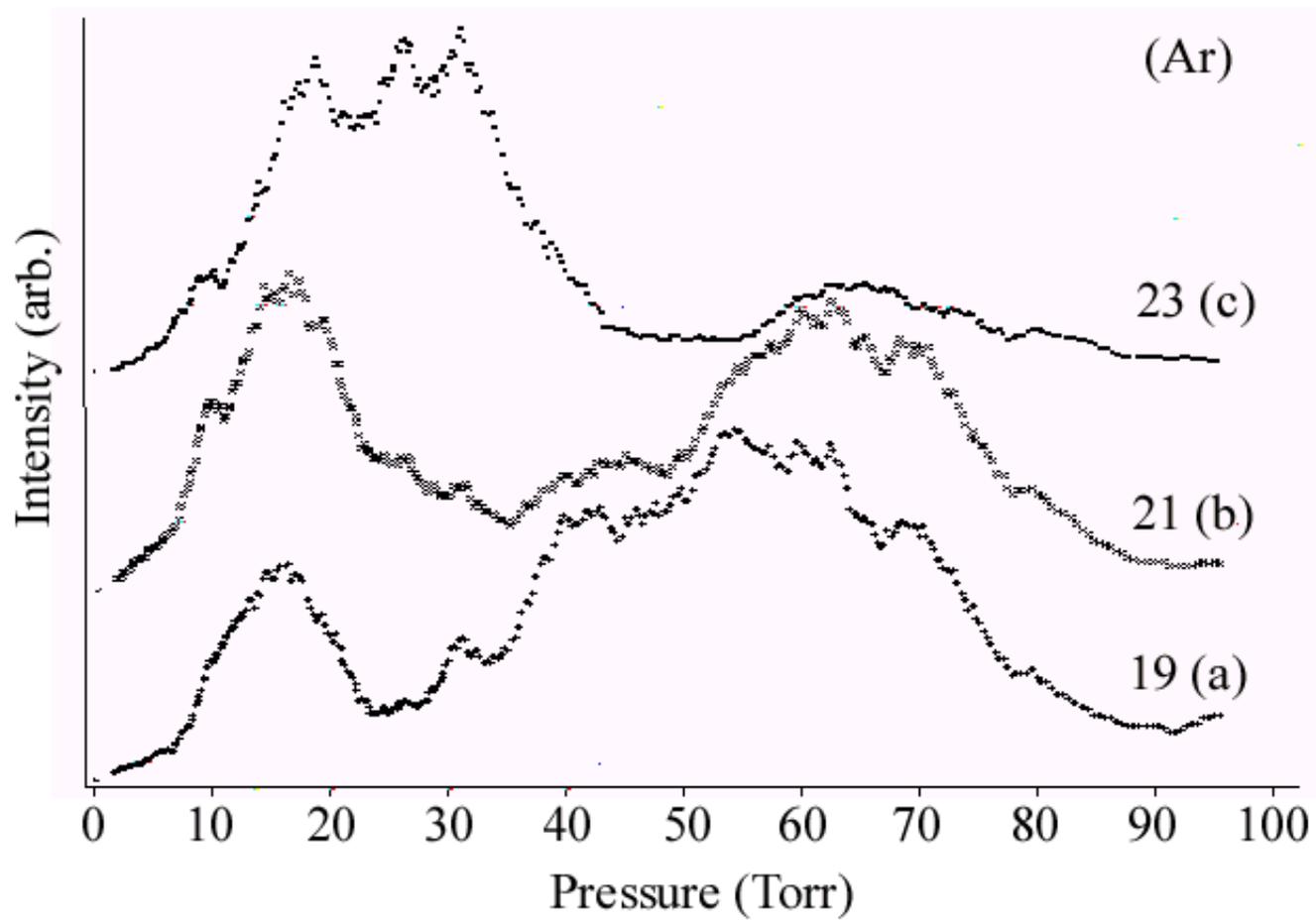

Figure 1d

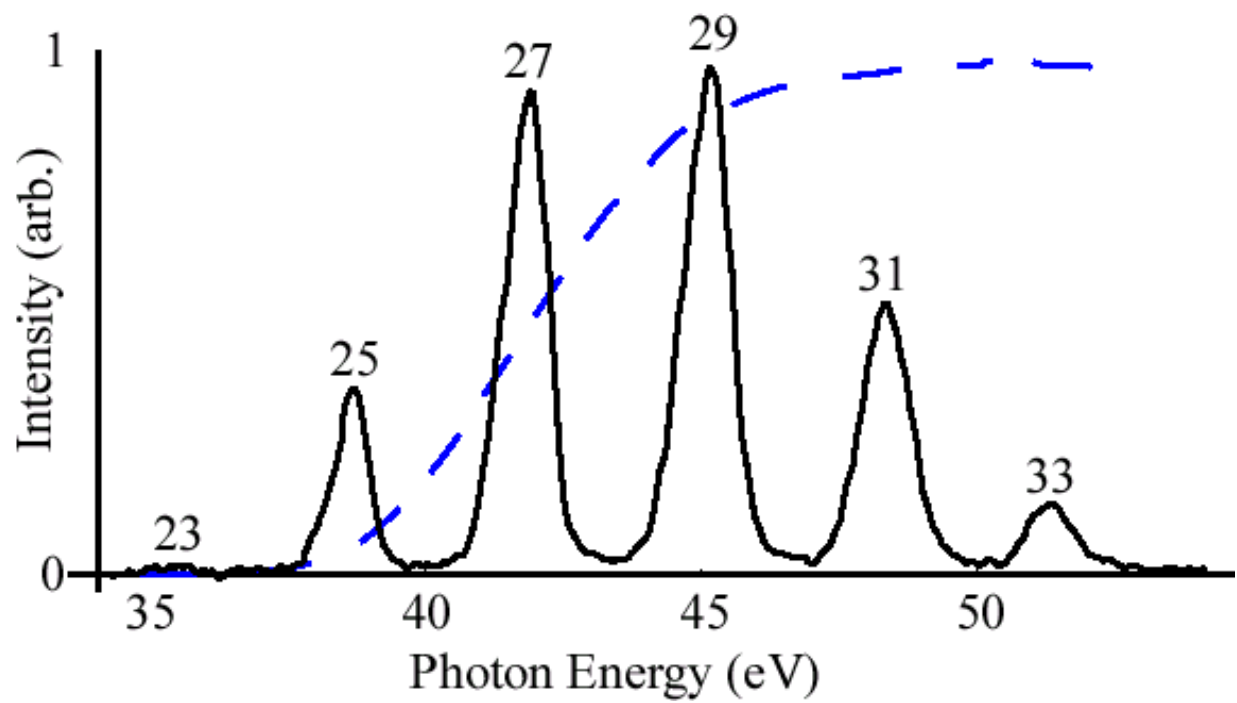

Figure 2

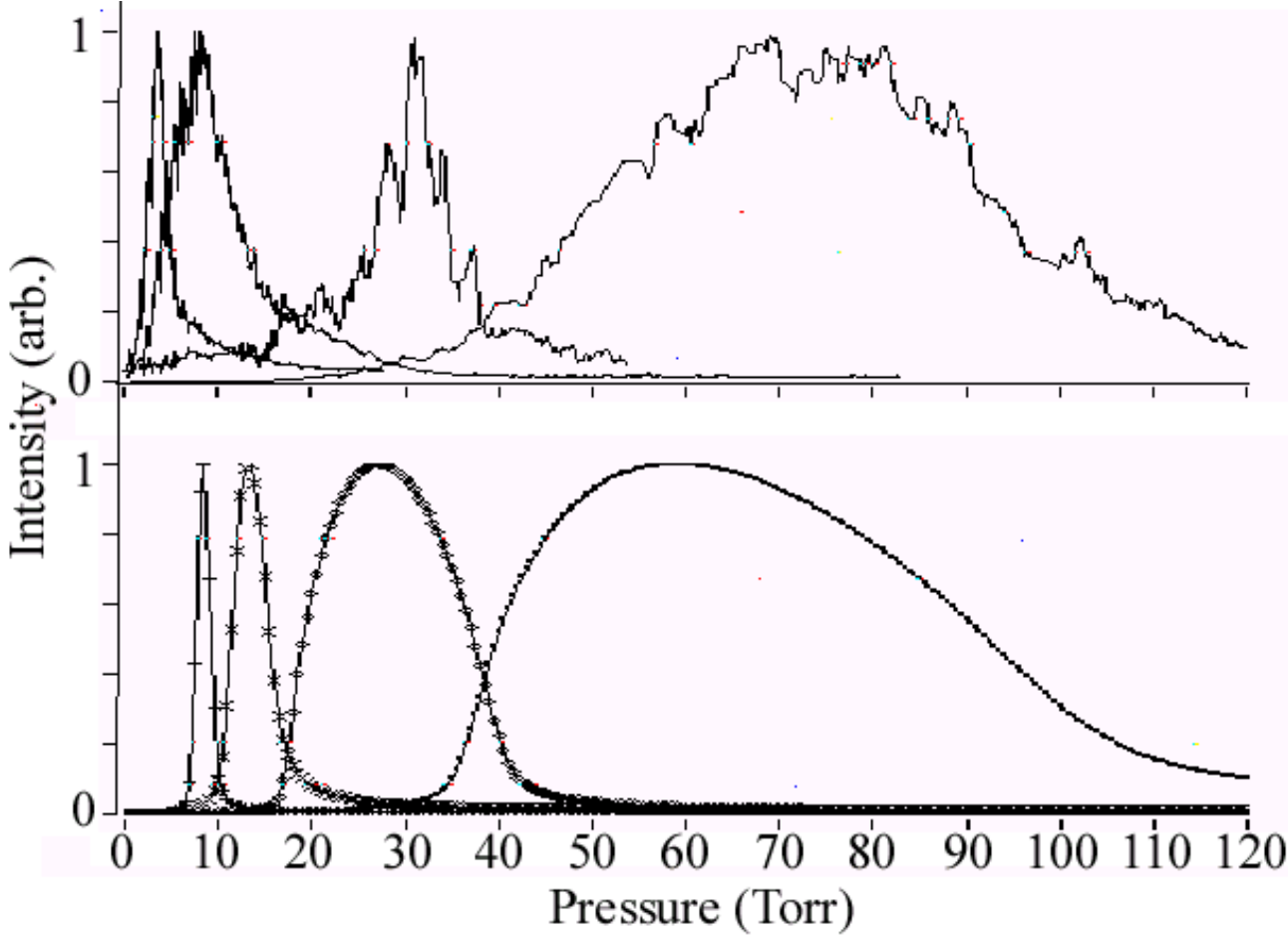

Figure 3a

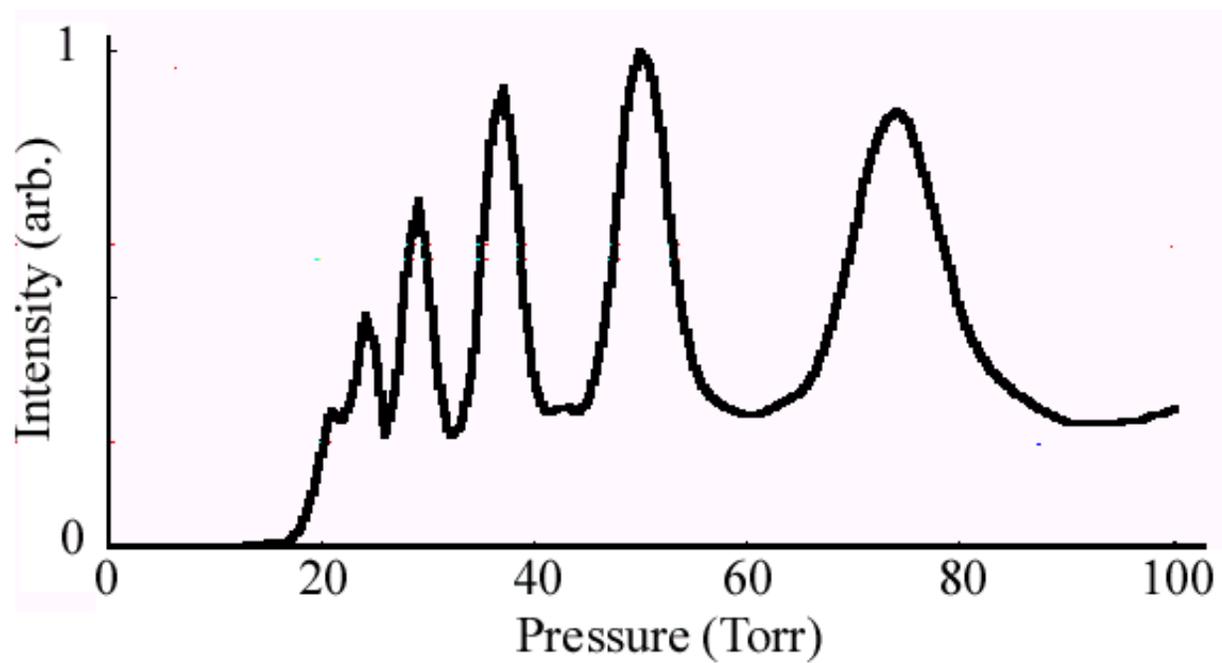

Figure 3b

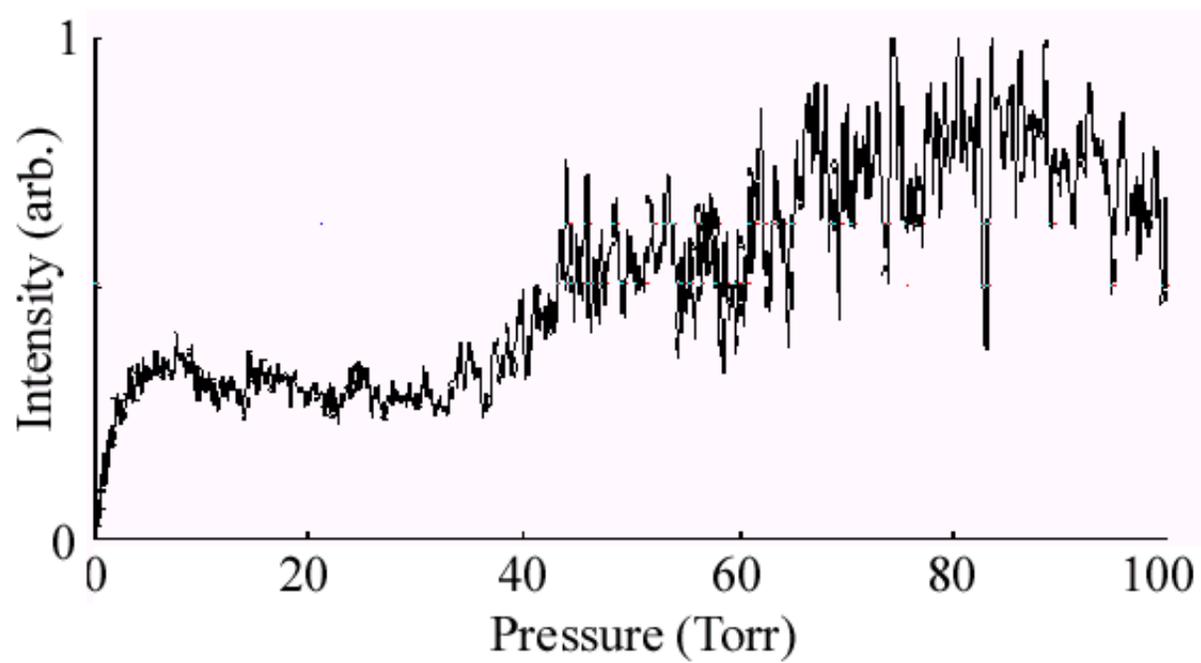

Figure 4

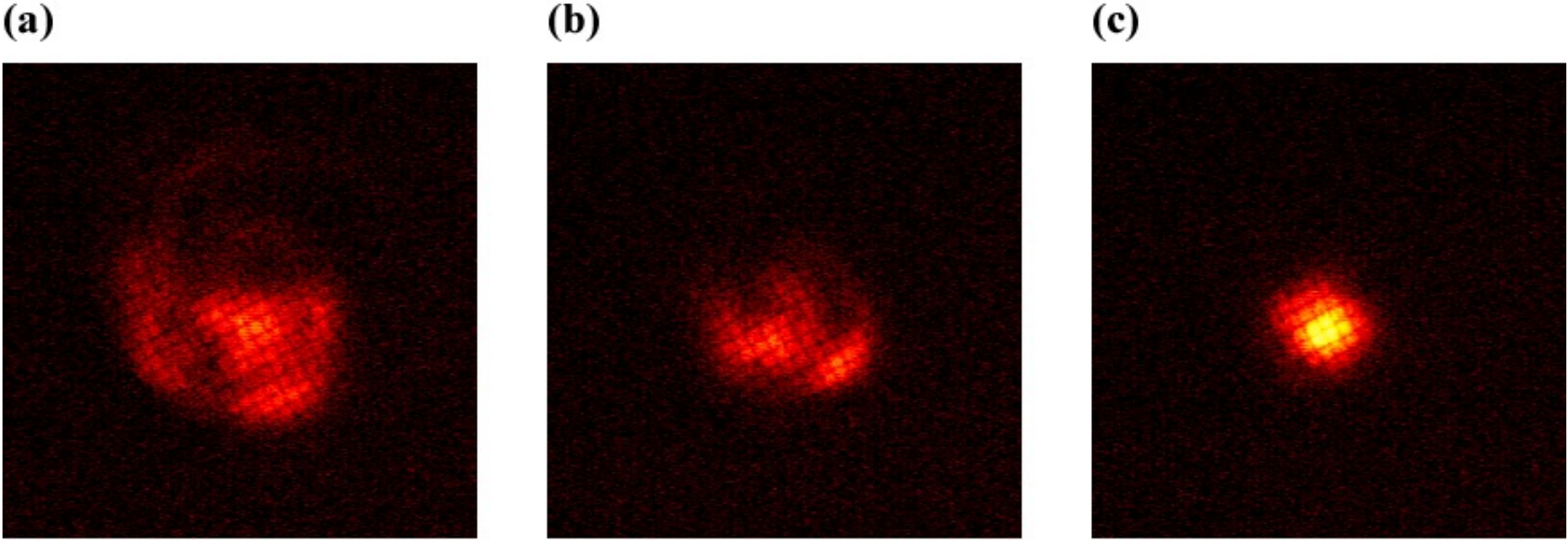